\documentclass[11pt,a4paper]{article}
\usepackage{amsmath,amssymb}
\usepackage{epsfig,graphicx}

\begin{document}

\title{\bf Astronomy meets QCD: \\
cooling constraints for the theories of internal
structure of compact objects}
\author{S.~B.~Popov$^a$\footnote{{\bf e-mail}:
polar@sai.msu.ru.},
D.~Blaschke$^{b,c}$\footnote{{\bf e-mail}: blaschke@theory.gsi.de},
H.~Grigorian$^{b,d}$\footnote{{\bf e-mail}: hovik.grigorian@uni-rostock.de},
B.~Posselt$^{e}$\footnote{{\bf e-mail}: posselt@mpe.mpg.de}
\\
$^a$ \small{\em Sternberg Astronomical Institute} \\
\small{\em Universitetski pr. 13, 119992 Moscow, Russia}\\
$^b$ \small{\em Institut f\"ur Physik, Universit\"at 
Rostock} \\
\small{\em D-18051 Rostock, Germany}\\
$^c$ \small{\em Bogoliubov Laboratory for Theoretical 
Physics, JINR} \\
\small{\em 141980 Dubna, Russia}\\
$^d$ \small{\em Department of Physics, Yerevan State 
University} \\
\small{\em 375049 Yerevan, Armenia}\\
$^e$ \small{\em Max-Planck-Institut f\"{u}r 
extraterrestrische Physik} \\
\small{\em D-85741 Garching, Germany}
}
\date{}
\maketitle

\begin{abstract}
We discuss a set of tests which confront observations of cooling compact
objects and theories of their thermal evolution.
As an example we apply the recently developed
$\mathrm{Log\, N}$-$\mathrm{Log\, S}$ test 
of compact star cooling theories 
to hybrid stars with a color superconducting quark matter core, we also
apply and discuss other existing tests. 
While there is not yet a microscopically founded superconducting quark matter 
phase which would fulfill constraints from cooling phenomenology, we explore
the hypothetical 2SC+X phase 
and show that the magnitude and density-dependence of the X-gap can be chosen 
to satisfy 
a set of tests: the temperature~--~age ($\mathrm{T}$-$\mathrm{t}$) test,  
the $\mathrm{Log\, N}$-$\mathrm{Log\, S}$ test, the brightness constraint,  
and the mass spectrum constraint.
Some recent modifications of the population synthesis model 
used to obtain the $\mathrm{Log\, N}$-$\mathrm{Log\, S}$ distribution
are briefly
discussed. In addition, we propose to use the age-distance diagram as a
new tool to study the local population of young isolated neutron stars.
\end{abstract}

\section{Introduction}

Compact objects known under the common name {\it neutron stars} belong to
the most facinating places in the Universe as the interest of different 
disciplines of science is focussed on them. 
In the present contribution we will elucidate how astronomical observations of 
cooling neutron stars (NSs) provide a tool to study the behaviour of matter
at high density. Here astronomy meets QCD.

Sometimes the term {\it neutron star} is used just as a common name for 
a large class of objects invented by the fantasy of theorists. 
They can be ``normal'' NSs, i.e. compact objects made mainly of hadrons even 
at the center -- {\it hadron stars}.
But there are other interesting possibilities. Central parts of NSs could 
contain more exotic forms of dense matter like a pion condensate or hyperonic
matter. Finally, as a result of the deconfinement transition hadrons get 
dissolved into their quark constituents due to the Pauli principle:
compact stars develop a quark matter core. 
In this case the NS is called a {\it hybrid star}. 
Eventually, the whole body (maybe except a tiny envelope) is converted into 
quark matter.   
In this case we speak of a {\it quark star}.
So, a large variety of interesting possibilities appears. 
One of the very few ways to test which particular scenario for the internal 
structure  is realized in Nature is to study the thermal evolution of compact 
objects. 
This paper is devoted to the problem of testing theories of NS cooling 
by confronting results of computations with observational data.

This contribution is mostly based on the paper \cite{pgb2006}.
Still, some new results and discussions are given.
At first, we describe different tests of cooling curves. 
Then we briefly describe the model for the cooling of hybrid stars 
to which we intend to apply the tests. 
In some details we dwell on the population synthesis scenario used here to 
produce the $\mathrm{Log\, N}$-$\mathrm{Log\, S}$ distributions.
After this we apply the test to a set of cooling curves of hybrid stars. 
Finally, we discuss further improvements in the scenario and present our
conclusions.

\section{Tests of cooling curves}

Testing the behaviour of matter in different regions of the QCD phase
diagram is an extremely important but difficult task. 
The region corresponding to  high density, but low temperature is not 
accessible to direct studies in terrestrial laboratory experiments. 
Therefore, astronomical data on the mass-radius relation of NSs,
on the gravitational redshift of spectral lines from these objects, 
etc., can be used together with
available data from ground-based experiments to get some constraints on the
properties of dense matter \cite{klahn,klahn2}.
Among astronomical observations those of cooling NSs are most important for 
studies about physical processes at the deconfinement phase transition in the 
QCD phase diagram. 
The idea is to compare calculated cooling curves with  data obtained from 
astronomical observations.
In this section we discuss different approaches to do this. 

\subsection{Age vs. temperature}
 
The most common test is the following. 
One just selects sources with relatively well known ages 
and temperatures and confronts data
points with theoretical cooling curves, see Fig.~1.
Naively it is assumed that if all data points can be covered by cooling
tracks then the model is considered being in correspondence with
observations.

The main advantages of this test from the point of view of its 
use by the community are the following two:

\noindent
1. It is clear and direct.

\noindent
2. Everybody who calculate the theoretical curves can do it as
observational data are available in the literature.

The test is widely used and was very well described many times (see, for
example, \cite{page2004,Blaschke:2004vq} and references therein). 
So, we do not give many details. 
Let us just specify few disadvantages which can be overcome if one
uses additional tests and considerations.

A. Well determined temperatures and, especially, ages are known for very
few objects. So, statistics is not very large.

B. Even if age and temperature estimates are available, they have
significant uncertainties, or they depend on the chosen model.

C. Objects with known temperature and ages form a very non-uniform sample, as
they were discovered by different methods with different instruments.
Different selection effects are in the game.

D. Mostly, the objects used in the test are younger than $10^5$ years.

E. There are some additional pieces of data which are not used in the
analysis (like the knowledge about mass distribution, non-detection of some
kinds of sources, etc.).

In the following subsections we focus on several additional methods which can
help to improve the situation when confronting theory with observations.

\begin{figure}[h]
\includegraphics[width=0.9\textwidth,angle=-90]{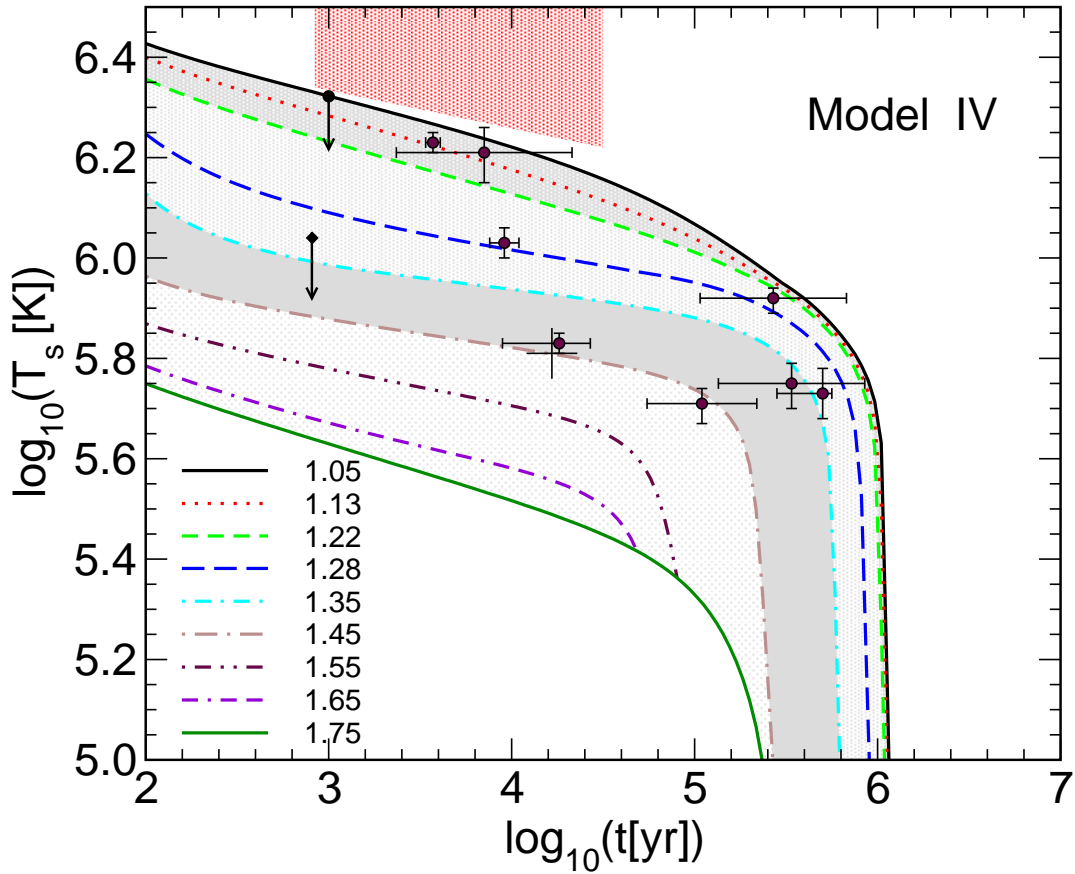}
\caption[]{ Hybrid star cooling curves for Model IV from \cite {pgb2006}. 
Different lines correspond to compact star mass values indicated in
the legend
(in units of $M_\odot$), data points with error bars are taken from 
\cite{page2004}.
The shaded areas corresponds to different bins in the mass spectrum shown in
Fig.~2.}
\label{fig:bc4}
\end{figure}

\subsection{Log N -- Log S}
The Log N -- Log S diagram is a useful tool in astrophysics.
Here N represents the number of sources with observed 
fluxes (at some energy range) larger than S. 
So, this is an integral distribution, i.e. it always grows towards
lower fluxes. 

This tool was used several times to study properties of the isolated NSs
population. In \cite{pgtb2006} we proposed to use the Log N -- Log S diagram
as an additional test of cooling curves of NSs.
The idea is to compare the observed Log N  -- Log S distribution with the
calculated in the framework of population synthesis approach (see below),
and to extract from this comparison a statement whether the model fits the 
data.

Our reasoning in favour of the new test is the following:

1. Thanks to the observations made onboard the ROSAT X-ray satellite we
have a relatively  uniform sample of NSs with detected thermal radiation.

2. The test doesn't require the knowledge of ages, temperature, etc.
    Only fluxes (which are well determined) and numbers are necessary to use
this test.

3. The test is sensitive to older ($\sim 1$~Myr) sources.

4. Most of ingredients of the population synthesis scenario except the cooling
curves can be fixed relatively well.

In \cite{pgtb2006} we were able to demonstrate that the tool really works
well as it puts additional constraints to the standard temperature vs. age 
test.

One of 
the main disadvantages of the test is that one needs to have a computer code 
to test a set of cooling curves. 
The way out can be to develope a web-tool
where everyone can download cooling curves and obtain the Log N -- Log S
distribution for selected parameters of the scenario. 
We hope to provide such a resource in future.
Another disadvantage is related to the precision of a population synthesis 
model.
Not all ingredients are equally well known, and a big piece of astrophysical
work has to be done to produce a good model. However, we believe that for 
young objects in the solar vicinity this problem can be solved, and in our
models we are on the right way to do it.

\subsection{Brightness constraint}

  Still, it is interesting to see how the temperature vs. age 
(T~--~t) test
can be
modified, and if it can help to select models successfully without the Log N
-- Log S test. Such a modification can be very helpful  
as in order to use the latter test 
it is necessary to apply  some complicated computer code which is not
publicly available at the moment. 
In this and in the following subsections we discuss two possible
additional tests.

In \cite{g2006} it was proposed to take into account the fact
that despite many observational efforts very hot NSs (log~T~$>6.3$~-~6.4~K)
with ages $\sim
10^3$~--~$10^4$~yrs were not discovered. If they would
exist in the Galaxy, then 
it would be 
very easy to find them (unless the interstellar absorption prevents us
to see a source, but absorption is not equally important in all directions:
so there are relatively wide ``windows'' to observe a significant part of
the Galaxy). 
If we do not see any very hot NSs, then we have to conclude
that at least their fraction is very small. 
This means, that any model
pretending to be realistic should not produce NSs with realistic masses with
temperatures higher than the observed ones. The region of avoidance can be
found in the Fig.~1: this is the hatched trapezoidal area on the top of all
curves.    

 This constraint is very sensitive to the properties of the crust of a NS
(see \cite{g2006} for details). 
Fitting the crust one can usually find a solution to satisfy the brightness 
constraint. 
So, this technique is an
useful addition to the standard temperature vs. age test.
The usage of only T~--~t plus the
brightness constraint method can lead to a wrong solution as both are not
very sensitive to the behaviour of the cooling curves for ages larger $\sim$
few $\times 10^5$ years, and just fitting the crust can result in  a
solution which can be proven wrong on the basis of the  
Log N -- Log S test when
properties of the internal parts of a NS are not properly selected. 
So, it is important to remember that the Log N -- Log S test is not very
sensitive to the crust properties. 
We conclude, that anyway the Log N -- Log S test should be used, too,
because
such complex approach helps to make a more complete testing of cooling
curves.

\subsection{Mass spectrum constraint}

 The mass spectrum of NSs is an important ingredient of the population synthesis
scenario. Normally, if we consider masses in the range $1\,M_\odot < M <
2\,M_\odot$, lighter stars cool slower. 
Our estimates of the mass spectrum \cite{ptpct2005,pgtb2006,pgb2006} show
that the fraction of NSs with masses larger than
$\sim 1.4\, M_\odot$ is very small. This means that more massive objects
should not be used to explain observations, especially if we speak about
bright or/and typical sources. In particular, close-by young NSs, like Vela,
should not be explained as massive stars as this is very improbable that we
are lucky to have such a young object (age $\sim$ 10 000 years) so close.
As we show in \cite{pgb2006} this simple constraint helps to close some
models which can successfully pass T-t or/and Log N -- Log S tests.

In Fig.~2 darker bins corresponds to more abundant NSs. The same colors are
used in Fig.~1.
So, ideally, on the
T--t plot most of data points should fall inside darker regions. As one can
see in Fig.~1, for Model IV this is the case. 

\begin{figure}[h]
\includegraphics[width=0.95\textwidth,angle=-90]{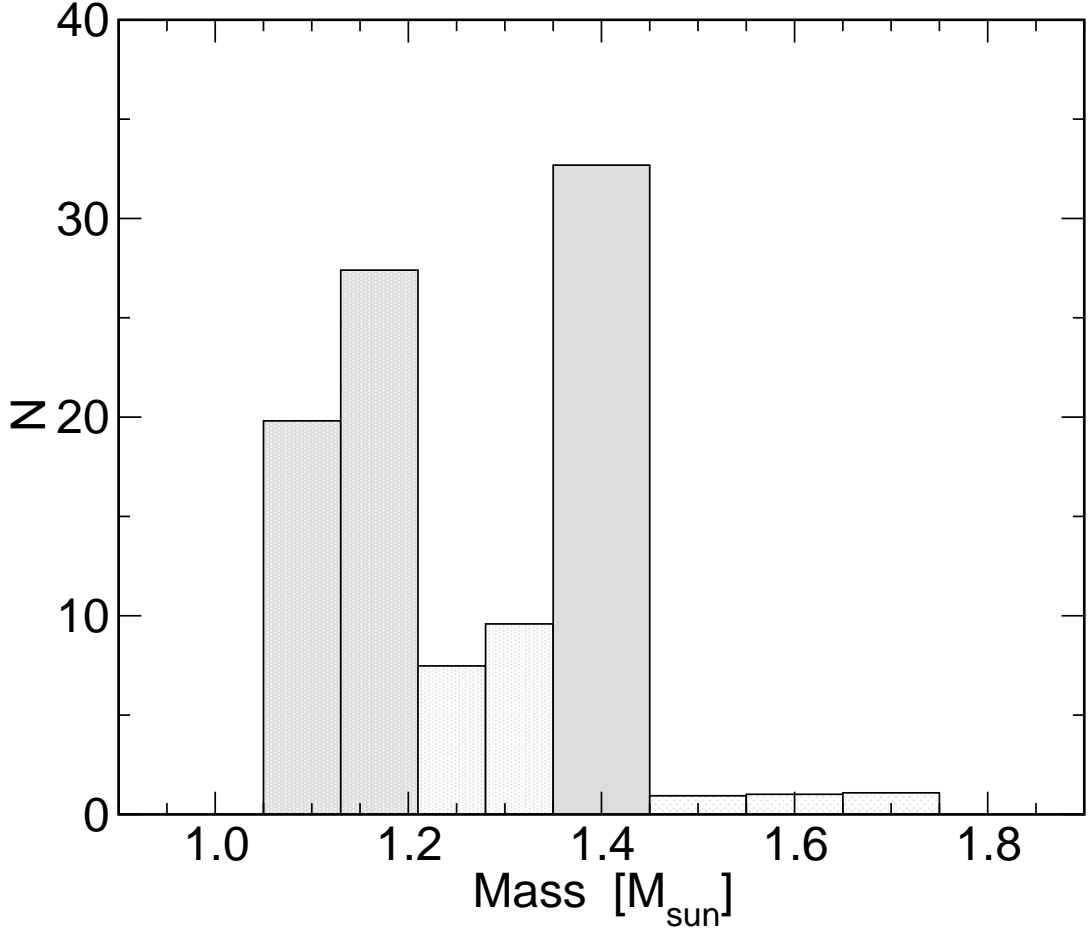}
\caption[]{The adopted mass spectrum, binned over eight intervals of
different widths. 
The grey tones for the mass bins encode their abundance and correspond to
those used in Fig.~1. Darker bins corresponds to more abundant NSs.
On the vertical axis the percentage of NSs in each bin is given.
}
\label{fig:mass}
\end{figure}

\section{Cooling model for hybrid stars} 
The details of calculations of the cooling curves used here
can be found in \cite{gbv2005}. We just briefly outline the main issues.

In the hadronic part of a hybrid star 
the main processes are the medium modified Urca  and the pair breaking
and formation  processes for our adopted equation of state  
of hadronic matter.  The possibilities of pion condensation and of other so
called
exotic processes in hadronic matter may be disregarded since these processes
have threshold densities at or above the critical
density  for the occurence of quark matter. 
For  the calculation of the cooling of the quark core in the hybrid
star we incorporate the most
efficient processes: the quark direct Urca  processes on
unpaired quarks, the quark modified Urca, the quark
bremsstrahlung, the electron bremsstrahlung, and the
massive gluon-photon decay. We include the emissivity of the quark pair
formation and breaking
processes. The specific heat incorporates the quark
contribution, the electron contribution and the massless and
massive gluon-photon contributions. The heat conductivity contains
quark, electron and gluon terms.

The 2SC phase has one unpaired color of quarks (say blue) for which
the very effective quark direct Urca process works and leads to a too fast  
cooling of the hybrid star in disagreement with the data.  We have suggested
to assume a weak pairing channel
which could lead to a small residual pairing of the hitherto
unpaired blue quarks. We call the resulting gap $\Delta_X$ and
show that for a density dependent ansatz:

\begin{equation}
\Delta_{\mathrm{X}}= \Delta_0 \, \exp{\left[-\alpha\, \left(
\frac{\mu - \mu_c}{\mu_c}\right)\right]}
\label{gap}
\end{equation}
with $\mu$ being the quark chemical potential, $\mu_c=330$ MeV.
Here we use different values of $\alpha$ and $\Delta_0$.

The physical origin of the X-gap remains to be identified. It
could occur, e.g., due to quantum fluctuations of color neutral
quark sextett complexes. Such calculations have not
yet been performed with the relativistic chiral quark models.

\section{Population synthesis model}

Population synthesis is a frequently used technique in astrophysics.
It is described, for example, in the review \cite{pp2005} where further   
references can be found.
The idea is to construct an evolutionary scenario for an artificial
population of some astronomical objects. The comparison with observations
gives the opportunity to test our understanding of evolutionary laws and 
initial conditions for these sources.

 We use the advanced version of the population synthesis scenario introduced
in \cite{pcptt2003}.   The main ingredients of the population synthesis
model we use are:
\begin{itemize}
\item the initial
distribution of NSs and their birth rate; 
\item the velocity distribution of NSs;
\item the mass spectrum of NSs; 
\item cooling curves; 
\item interstellar absorption;
\item properties of the detector.
\end{itemize} 
We assume that NSs are born in the Galactic disc and in the Gould Belt.
The disc region is calculated up to 3 kpc from the Sun, and is assumed to be
of zero thickness. The birth rate in the disc part of the distribution
is taken as 250 NS per Myr. 
The Gould Belt is modeled as a flat disc-like structure with a hole
in the center. The inclination of the Belt relative to the galactic plane is
$18^{\circ}$.  The NS birth rate in the Belt is 20 per Myr.  

The population synthesis code calculates spatial trajectories of NSs
with the time step $10^4$~yrs. For each point from the set of cooling curves
we have the surface temperature of the NS. 
Calculations for an individual track are stopped at the age
when the hottest  NSs reaches the temperature $10^5$~K.

As we known the surface temperature, we can calculate the radiation
flux in the energy band of interest. 
In principle, it is possible to take into account atmospheric properties of
NSs, as even a tiny atmosphere can influence the spectrum of an object.
However, there are many unsolved issues here. A large variety of atmospheric
models is discussed, and it is impossible to make a choice in favor of one
particular model. 
On the other hand, for many sources a blackbody fit
gives very good results, and so
all our calculations are done for  pure blackbody spectra. 

With the known distance from the Sun and the ISM
distribution we calculate the column density, which is necessary to derive
absorption as  soft X-rays fluxes are significantly absorbed. 
Finally, on the base of known unabsorbed flux and column density 
count rates are calculated using the ROSAT response matrix. 
Results are summarized along
each individual trajectory. We calculate 5,000 tracks for each model.   
The results are then normalized to the chosen NS formation rate
(270 newborn NSs per Myr in the whole region of the considered problem).  

\section{Results of tests when applied to cooling curves of hybrid stars}

In this section we present our results for the best model (IV) studied in
\cite{pgb2006}.  The calculated Log N -- Log S distributions for different
values of parameters are shown in Fig.~3.
We used two different values for the size of the Gould Belt (300 and 500
pc), and used two variants of the mass spectrum: the full one and truncated.
In Fig.~2 the full spectrum is shown. For the truncated one the contributions of
the first two mass bins are added to the third one. The reason for such a
modification is related to the fact that it is not clear whether a significant
amount of light NSs   is formed \cite{tww96}. 

\begin{figure}[h]
\includegraphics[width=0.9\textwidth,angle=-90]{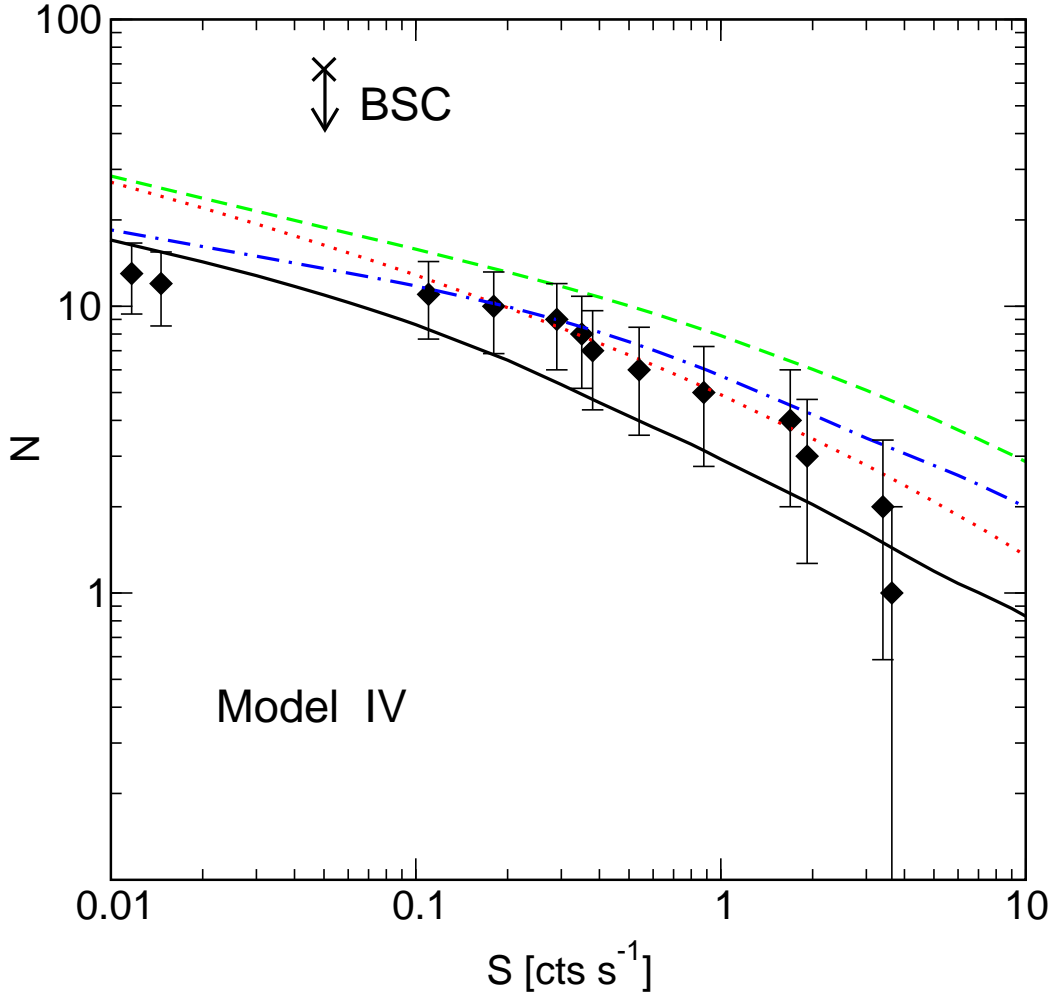}
\caption[]{
$\mathrm{Log\, N}$-$\mathrm{Log\, S}$ distribution for Model IV.
Four variants are shown:
$\mathrm{R_{belt}}=500$~pc and truncated mass spectrum (full line),
$\mathrm{R_{belt}}=500$~pc and non-truncated mass spectrum (dotted line), 
$\mathrm{R_{belt}}=300$~pc and truncated mass spectrum (dash-dotted line),
and finally $\mathrm{R_{belt}}=300$~pc (dashed line)
for non-truncated mass distribution. 
BSC corresponds to the limit obtained in \cite{bsc}.}
\label{fig:m4}
\end{figure}

This model succesfully passes all four test (T-t, Log N -- Log S, mass
constraint, brightness constraint). Other three models studied in
\cite{pgb2006} failed to pass one (or more) of the four tests.

\section{Modifications of the population synthesis scenario}

 Though, the main parameters of the population synthesis scenario for
close-by young NSs are known, and we demonstrated that the scenario itself 
works well, we continue to upgrade the code. Below we briefly discuss the
main recent updates made in the model. Some details can  also be found in
\cite{ppht2006}.

\subsection{Initial spatial distribution}

 The first thing that can be improved in the model is the initial spatial
distribution of NSs, or, better say, distribution of progenitors.
Instead of using flat distributions in the galactic disc and in the Belt one
can use a realistic distribution of massive stars in the solar vicinity.

\begin{figure}[h]
\includegraphics[width=\textwidth]{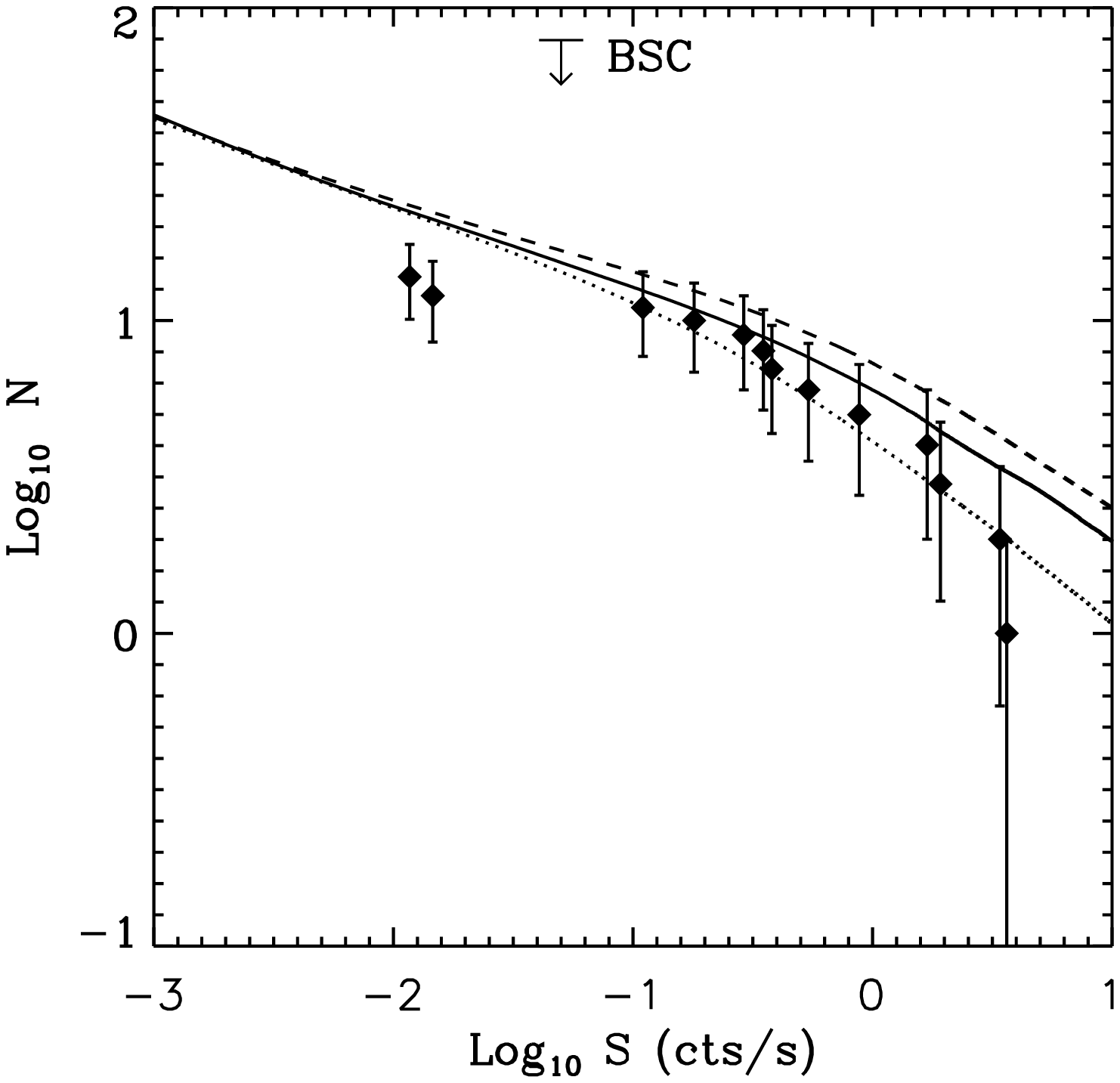}
\caption[]{
Three curves correspond to different initial distributions.
The new distribution (solid curve), and older
simple model for two different Belt radii (dashed and dotted curves).
The upper curve is for the Belt radius 300 pc, the lower -- for 500 pc.
BSC corresponds to the limit obtained in \cite{bsc}.
All curves are plotted for ``old'' (Fig.~2) truncated mass spectrum.
}
\label{fig:lnls1}
\end{figure}

In our upgraded model we construct the distribution of progenitors from three
parts. Inside 400 pc around the Sun we use actual data on the massive stars
distribution obtained by the Hipparcos satellite. 25 out of 270 NS per Myr
are assumed to come from
this population.  
Outside this region of up to 3 kpc we  use the data on associations of massive
stars. Data on 36 most populated associations is used \cite{ppht2006}.
245 out of 270 NS per Myr   
are assumed to come from
associations. The remaining 20 NS per Myr are randomly distributed in the plane
of the galactic disc between 0.4 and 3 kpc from the Sun, this population
represents ``field stars'' and stars from associations not included 
explicitly into our model.  

Comparison between old and new approaches can be seen in Fig.~4.
Three Log N --Log S distributions are calculated for the same cooling curves
(Model IV). Two of them are for the old simple model for two different Belt
radii, and the third one is calculated for the new initial distribution
model. 

\subsection{Mass spectrum modifications}

The mass spectrum of newborn NSs is not well known. 
A few observed sources which can
potentially help to solve this puzzle are binary radio pulsars, but it is
not clear if the mass spectrum for these objects can be applied for the
population of isolated (and often radio quiet) NSs. So, we can just try to
make a model of a spectrum which is 
based on available data on progenitors properties
and on calculations of stellar evolution and
supernova explosions. As computed stellar and supernova models are not very
precise one have to try to apply different results of such calculations.
 In addition to the ``old'' mass spectrum shown in Fig.~2 we tried a
modification.
 The main difference with the ``old'' one is that for progenitors with
initial masses $>12\, M_{\odot}$ we use the relation between progenitor mass 
and NS mass taken from \cite{hws2004}.
The full ``new'' mass spectrum is shown in Fig.~5.
In Fig.~6 we compare Log N -- Log S curves for different mass
spectra. For an illustration we add a curve for the flat mass spectrum
used in \cite{pcptt2003}.
According to this spectrum all masses are equally probable.
All these calculations are done for the new initial spatial distribution
of progenitors.

\begin{figure}[h]
\includegraphics[width=\textwidth]{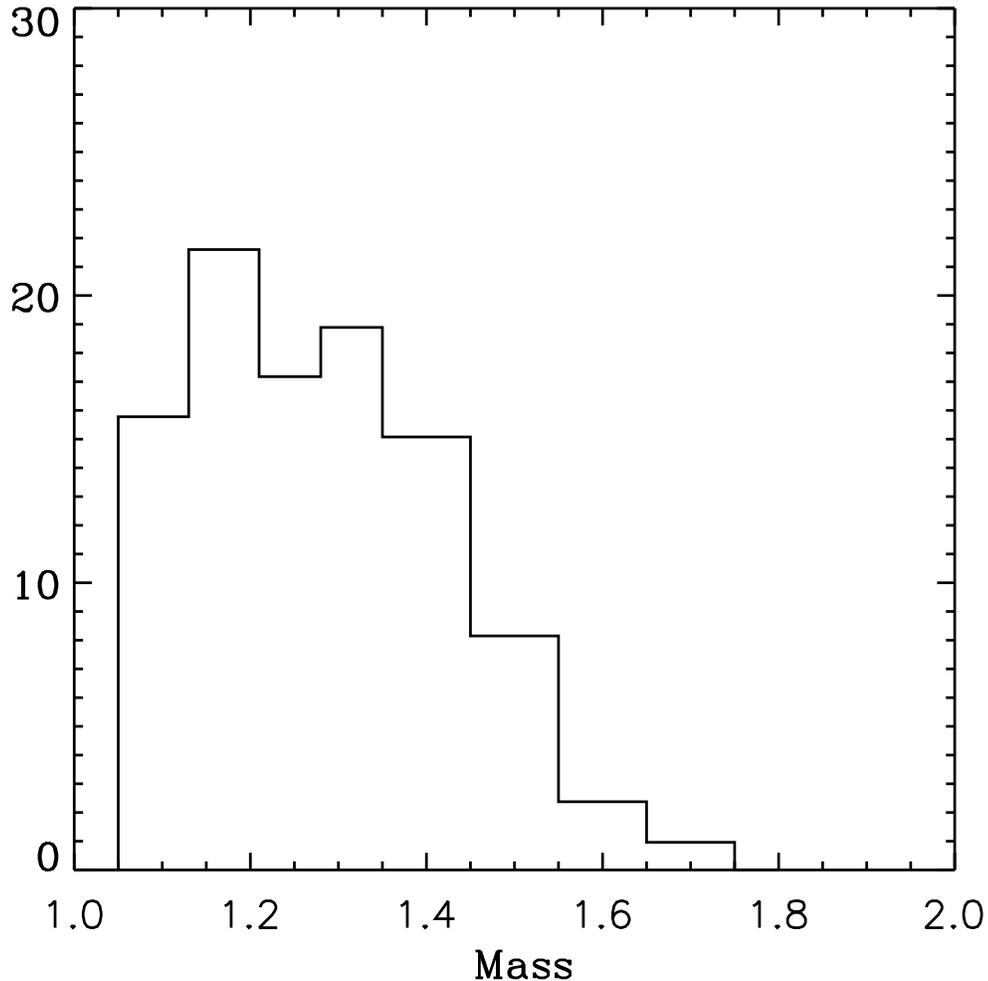}
\caption[]{The ``new'' mass spectrum, binned over eight intervals of
different widths. Gravitational is given mass in solar masses.
}
\label{fig:massnew}
\end{figure}

\begin{figure}[h]
\includegraphics[width=\textwidth]{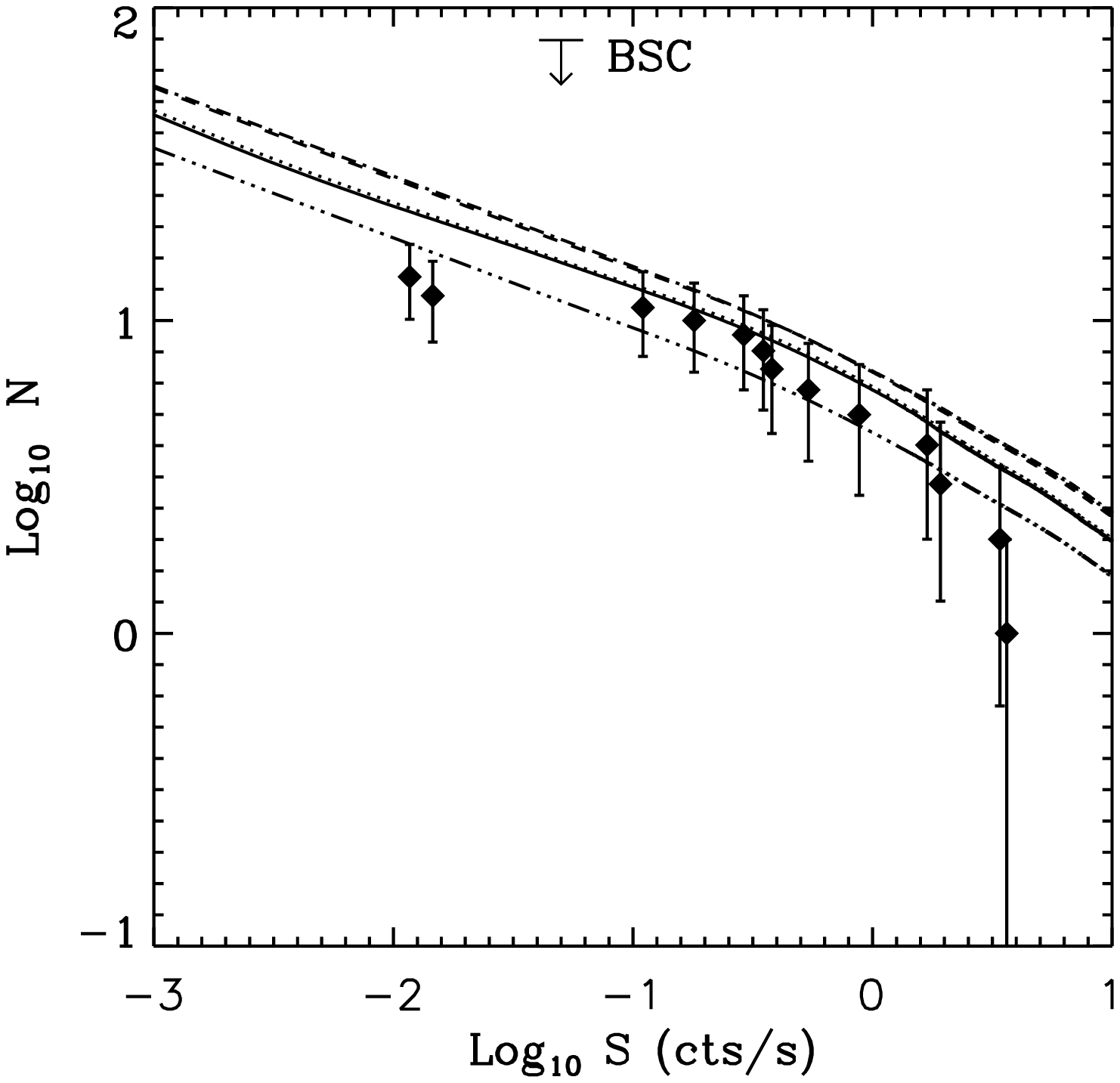}
\caption[]{
Two upper curves correspond to non-truncated mass spectra
(dashed to the ``new'' and dot-dashed -- to the ``old'').
Two curves in the middle -- to truncated spectra (solid --
``old'', dotted -- ``new'').
The lowest curve is shown for the flat mass spectrum.
}
\label{fig:lnls2}
\end{figure}

Surprisingly, the new mass spectrum (Fig.~5) does not lead to serious
modifications of the Log N -- Log S distribution.
Still, one has to be warned that further investigations of the mass spectrum
of newborn isolated NSs are necessary. In particular, it is important to
understand how mass can be correlated with other initial parameters of NSs.

\subsection{Further improvements}

 Several other important improvement were made recently in our population
synthesis model. In particular, a realistic distribution of the interstellar
medium was taken into account. The procedure to calculate the count rate
for the ROSAT devices was updated. Finally, we added the possibility to
compute the Log N -- Log S distribution for the XMM-Newton satellite.
These changes will be described elsewhere.

\subsection{Age-Distance diagram}
\begin{figure}[h]
\includegraphics[width=\textwidth]{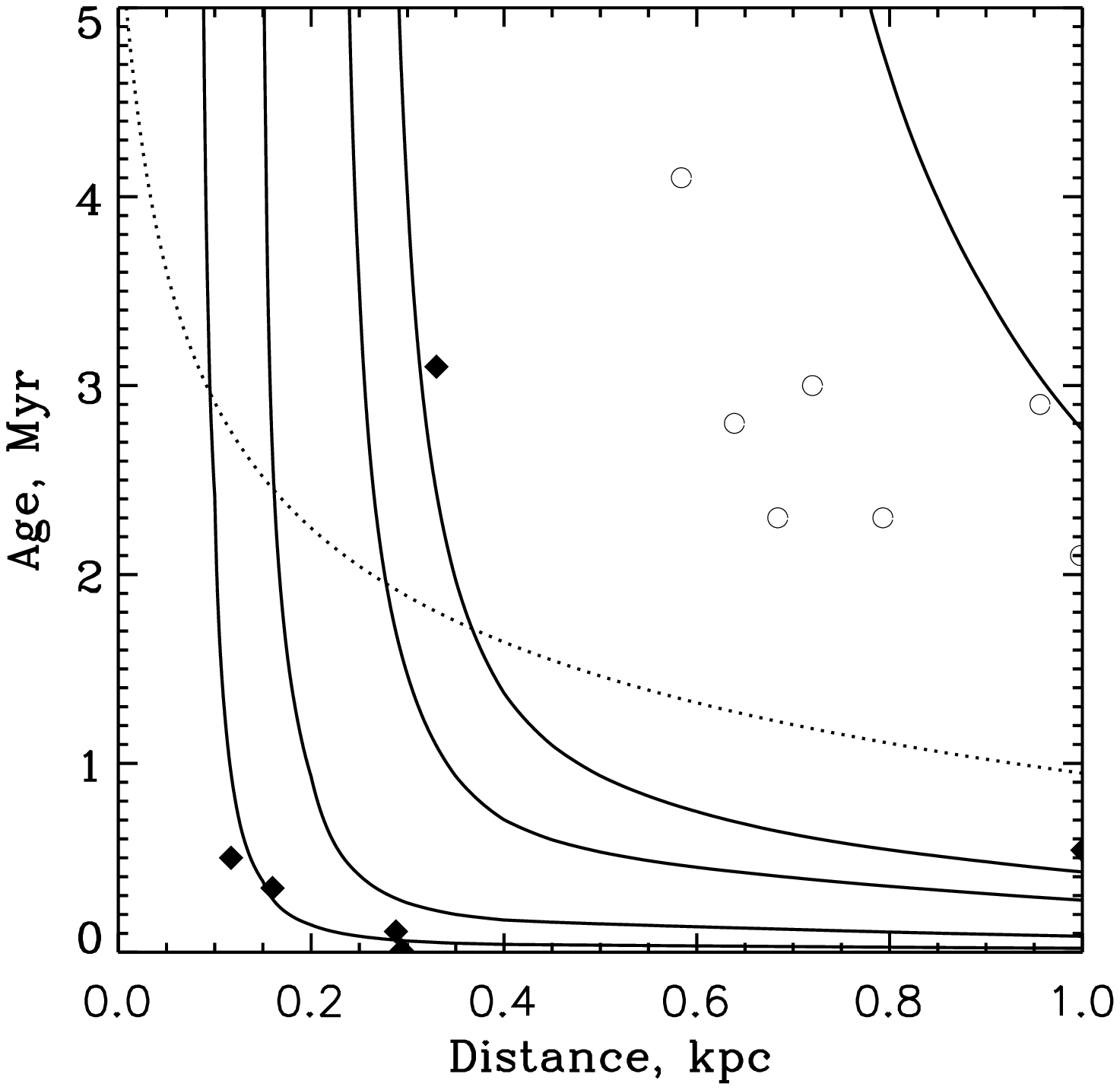}
\caption[]{Age-distance diagram for close-by young NSs with realistic
birth rate and initial distribution, 
dynamical effects are taken into account. 
Filled symbols -- sources detected in X-rays,
empty symbols -- 
known close-by young NSs (radio pulsars) with
non-detected  thermal emission. 
}
\label{fig:add}
\end{figure}

 In \cite{p2004} we introduced the age-distance diagram (ADD) for close-by
young isolated NSs. 
 This diagram, 
as it is clear from its name, represent a plot, where each point
corresponds to a star with known age and distance. In the Fig.~\ref{fig:add}
we show an example from \cite{p2004}.  

Five solid 
lines in Fig.~7 are plotted for 1, 4 
13, 20 and 100 sources.
Black and empty symbols represents known NSs (of course, as not for all of
them ages
are known, just a part of the population can be plotted in the graph).
The dotted line represents the ``visibility'' line.
This line shows a maximal distance for a given age (or vice versa
a maximal age for a given distance) at which a hot (i.e. low-mass) NS can be
detected. Here the line is plotted for a very simple model.
We just assume some limiting value of unabsorbed flux: 
$10^{-12}$~erg~cm$^{-2}$~s$^{-1}$. 
According to 
WebPIMMS\footnote{http://heasarc.gsfc.nasa.gov/docs/corp/tools.html}
it corresponds to $\sim0.01$ 
ROSAT PSPC counts per second 
for $N_H=10^{21}$~cm$^{-2}$ and a blackbody spectrum with $T=90$~eV,
or to $\sim0.1$ 
ROSAT PSPC counts per second
for $N_H=10^{20}$~cm$^{-2}$ and a blackbody spectrum with $T=50$~eV.
The latter values corresponds to the dimmest source among  the
Magnificent seven -- RX J0420.0-5022;
the former to possibly detectable hot distant objects.
Without any doubt such a simple approach
underestimates absorption at large distances. 
If used for testing cooling curves this line has to be calculated more
accurately.

 Unfortunately, at the present time ages and distances are known with enough
precision only for few young NS in our neighborhood.
 That is why at the
moment the ADD can be just of limited utility.  
But it seems, that as the data on close-by isolated NSs grows,
the ADD can become an additional tool to study this group of sources,
especially with population synthesis codes. 

The ADD potentially can be used as an additional test for cooling curves. 
The idea would be to model with a population synthesis code an
artificial population and compare it with the observed one using the ADD.
When ages and distances for all sources from the Magnificent seven and other
close-by young NSs are known
it can be done.
The advantage of this diagram is that one can add NSs which are unobservable
in thermal X-rays. 
Dark (in soft X-rays) NSs can (and should) also be used to test cooling
curves in the framework of population synthesis.

\section{Conclusions}

 In conclusion, we want to underline that the idea of the conferences like
QUARKS-2006 is a very fruitful one, as these meetings
are the places where theoretical
physicists and astrophysicists can have joint discussions: QCD meets
astronomy again.
Now it is obvious that there are theoretical models
which can be tested only in ``celestial laboratories''. Neutron stars
definitely are such labs. 

Observations of cooling NSs provide a rare
opportunity to study particular parts of the QCD phase diagram. 
We hope that the development of new tools to confront theory of cooling of
compact objects with observations can help to achieve progress in both: 
astrophysics and  theoretical physics.  

\bigskip

\noindent
{\bf Acknowledgments}
The work of SP is supported by the RFBR grant 06-02-16025, by the Cariplo
foundation and by the ``Dynasty'' Foundation. 
HG was supported in part by DFG under grant 436 ARM 17/3/05.
SP thanks the Organizers for partial support.

\end{document}